\documentclass[twoside,twocolumn,9pt]{article}
\usepackage{extsizes}
\usepackage[super,sort&compress,comma]{natbib} 
\usepackage[version=3]{mhchem}
\usepackage[left=1.5cm, right=1.5cm, top=1.785cm, bottom=2.0cm]{geometry}
\usepackage{balance}
\usepackage{mathptmx}
\usepackage{sectsty}
\usepackage{graphicx} 
\usepackage{lastpage}
\usepackage[format=plain,justification=justified,singlelinecheck=false,font={stretch=1.125,small,sf},labelfont=bf,labelsep=space]{caption}
\usepackage{float}
\usepackage{fancyhdr}
\usepackage{fnpos}
\usepackage[english]{babel}
\addto{\captionsenglish}{%
  
}
\usepackage{array}
\usepackage{droidsans}
\usepackage{charter}
\usepackage[T1]{fontenc}
\usepackage[usenames,dvipsnames]{xcolor}
\usepackage{setspace}
\usepackage[compact]{titlesec}
\usepackage{hyperref}

\definecolor{cream}{RGB}{222,217,201}

\begin{document}

\pagestyle{fancy}
\thispagestyle{plain}
\fancypagestyle{plain}{
%%%HEADER%%%
\renewcommand{\headrulewidth}{0pt}
}
\newcommand{\todo}[1]{\textbf{\textcolor{red}{#1}}}
\newcommand{\tx}{\text}
\newcommand{\nd}{\noindent}
\newcommand{\dd}{\text{d}}
%%%END OF HEADER%%%

%%%PAGE SETUP - Please do not change any commands within this section%%%
\makeFNbottom
\makeatletter
\renewcommand\LARGE{\@setfontsize\LARGE{15pt}{17}}
\renewcommand\Large{\@setfontsize\Large{12pt}{14}}
\renewcommand\large{\@setfontsize\large{10pt}{12}}
\renewcommand\footnotesize{\@setfontsize\footnotesize{7pt}{10}}
\makeatother

\renewcommand{\thefootnote}{\fnsymbol{footnote}}
\renewcommand\footnoterule{\vspace*{1pt}% 
\color{cream}\hrule width 3.5in height 0.4pt \color{black}\vspace*{5pt}} 
\setcounter{secnumdepth}{5}

\makeatletter 
\renewcommand\@biblabel[1]{#1}            
\renewcommand\@makefntext[1]% 
{\noindent\makebox[0pt][r]{\@thefnmark\,}#1}
\makeatother 
\renewcommand{\figurename}{\small{Fig.}~}
\sectionfont{\sffamily\Large}
\subsectionfont{\normalsize}
\subsubsectionfont{\bf}
\setstretch{1.125} %In particular, please do not alter this line.
\setlength{\skip\footins}{0.8cm}
\setlength{\footnotesep}{0.25cm}
\setlength{\jot}{10pt}
\titlespacing*{\section}{0pt}{4pt}{4pt}
\titlespacing*{\subsection}{0pt}{15pt}{1pt}
%%%END OF PAGE SETUP%%%

%%%FOOTER%%%
\fancyfoot{}

\fancyfoot[RO]{\footnotesize{\sffamily{1--\pageref{LastPage} ~\textbar  \hspace{2pt}\thepage}}}
\fancyfoot[LE]{\footnotesize{\sffamily{\thepage~\textbar\hspace{3.45cm} 1--\pageref{LastPage}}}}
\fancyhead{}
\renewcommand{\headrulewidth}{0pt} 
\renewcommand{\footrulewidth}{0pt}
\setlength{\arrayrulewidth}{1pt}
\setlength{\columnsep}{6.5mm}
\setlength\bibsep{1pt}
%%%END OF FOOTER%%%

%%%FIGURE SETUP - please do not change any commands within this section%%%
\makeatletter 
\newlength{\figrulesep} 
\setlength{\figrulesep}{0.5\textfloatsep} 

\newcommand{\topfigrule}{\vspace*{-1pt}% 
\noindent{\color{cream}\rule[-\figrulesep]{\columnwidth}{1.5pt}} }

\newcommand{\botfigrule}{\vspace*{-2pt}% 
\noindent{\color{cream}\rule[\figrulesep]{\columnwidth}{1.5pt}} }

\newcommand{\dblfigrule}{\vspace*{-1pt}% 
\noindent{\color{cream}\rule[-\figrulesep]{\textwidth}{1.5pt}} }

\makeatother
%%%END OF FIGURE SETUP%%%

%%%TITLE, AUTHORS AND ABSTRACT%%%
\twocolumn[

\vspace{1em}

\begin{tabular}{p{2cm} p{14cm} p{2cm}}
&\noindent\centering\Huge{\textbf{Self-propelled particles driven by light}}& \\

& \vspace{0.3cm} & \\

& \centering \large{Jannis Fischer
 $^{\ast}$\textit{$^{,\,a}$},  Alejandro Jurado%Jiménez
 \textit{$^{a}$}
 and Timo Betz\textit{$^{a}$}\textit{$^{,\,b}$}}& \\

& \vspace{0.8cm} & \\

&\normalsize{Recent advances in the field of active soft matter promise a lot. Both, experimental advances and theoretical understanding point towards new material classes in reach, for example self-healing materials that might switch their properties from elastic to solid easily or switch their macroscopic shapes. All these materials require an active force to propel parts of themselves on the micrometer scale. While chemical fuels are often used to generate these active forces, applying energy in a simple and continuous way remains unsolved. Here we explore using light as such an energy source. Overall, generating active driven, self-propelled particles is hence not only of great interest but also a general challenge. Moreover, controlling such particles even within living tissue would open new worlds, for example to enable specific drug delivery or the design of micro-robots. One recently proposed method to establish light driven self propelled particles is to create specific shaped and transparent objects, that move when illuminated with homogeneous light. In these particles, the refraction of the light leads to a momentum transfer, which then drives the active movement\cite{Jurado2018Photonic}. Here, we show both in simulation and experiments that the production of such particles is possible and demonstrate the feasibility of this propulsion effect, while investigating different shapes. Our experiments show that breaking the shape-symmetry of the particles creates a refraction-based propulsion under homogeneous illumination. Subsequent simulations reveal that total reflection leads to the largest momentum transfer among all different geometries considered. Overall, our study introduces the proof-of-principle for refraction-propelled particles, which has the potential to benefit many fields of study including cellular behaviour, collective dynamics and the understanding of disease mechanisms.
}&
\end{tabular}

\vspace{0.8cm}

]
%%%END OF TITLE, AUTHORS AND ABSTRACT%%%

%%%FONT SETUP - please do not change any commands within this section
\renewcommand*\rmdefault{bch}\normalfont\upshape
\rmfamily
\section*{}
\vspace{-1cm}

%%%FOOTNOTES%%%
\footnotetext{\textit{$^{*}$~ E-mail: jannis.fischer@phys.uni-goettingen.de}}
\footnotetext{\textit{$^{a}$~Third Institute of Physics, Georg-August-Universität Göttingen, Friedrich-Hund-Platz 1, 37077 Göttingen, Germany.}}

\footnotetext{\textit{$^{b}$~Cluster of Excellence ’Multiscale Bioimaging: From Molecular Machines to Networks of Excitable Cells’ (MBExC), Georg-August-Universität Göttingen , Universitätsmedizin Göttingen , Robert-Koch-Str. 40 , 37075 Göttingen , Germany}}

%%%END OF FOOTNOTES%%%

%%%MAIN TEXT%%%%

\section{Introduction}
In his famous lecture "There's plenty of room at the bottom" in 1959, Richard P. Feynman foresaw many of the technological challenges and discoveries that our curiosity for "miniaturization" has given rise to. The electron microscope has been improved since then, many biological questions regarding the behaviour of DNA and proteins have been sorted out, single atoms can be manipulated, and we are moving towards small computers "able to make their own judgements" \cite{feynman1992plenty}. However, one of his key comments was left mainly unaddressed: why don't we build more microscopic machines able to drive around and perform microscopic tasks? How come we have not yet be able to "swallow the surgeon" before a complicated heart intervention, who would be able to cut and repair blood vessels with  tiny surgical instruments?
The challenge holding us back at this point is not microfabrication anymore, but rather the control of movement at the micro-scale, where forces and inertia don't act in an intuitive manner. 
For the past decades, self-propelled particles (SPPs) have been one of the most promising ideas for the fabrication and control of microscopic agents. However, most of the approaches rely on an specific fuel-based propulsion, which make in-vivo biological applications not viable. In this study we present a proof-of-principle for light-based movement of small polymer particles, engineered to refract light in a preferential way to propel themselves under a homogeneous light field. In both, experiments and simulations, we show the feasibility of this simple, yet largely unexplored mechanism and we propose optimization ideas for future implementations.

\subsection{Reynolds number}

On the scale of single cells, forces act on objects in different ratios than we are familiar with from everyday life. A good starting point to understand the fundamental mechanical differences that arise at the microscale is to revise the concept of force. We know from our everyday experiences that inertial forces usually outweigh frictional forces: a moving car has a certain stopping distance when braking and when starting off on a bicycle it takes a few seconds to reach a comfortable speed. On small scales, however, these forces are distributed differently: The smaller an object is, the greater the role friction plays, up to the point where it is common to neglect inertial forces for very small objects \cite{Purcell_LifeLowReynoldsNumber}. 

While the inertia contribution scales with the third power of an object's size $L^3$, its friction scales linearly \cite{stokes_200938}:
\begin{equation}
    F_\tx{friction} = 6 \pi L \eta v \tx{ ,}
\end{equation} 
which leads to frictional forces playing a much a bigger role the smaller the object is. In the above equation, $\eta$ is the dynamic viscosity of a surrounding fluid and $v$ the velocity of the object. In concrete terms, this shows why large ships glide significantly longer with the engine switched off than small ones, even though their total friction is larger. To better compare the relative weight of inertial and frictional forces in moving objects, the Reynolds number, as described by Osborne Reynolds at the end of the 19th century \cite{Reynolds}, is usually used\cite{Fluid_Dynamics54}:

\begin{equation}
    \tx{Re} = \frac{\rho v L}{\eta} \tx{ ,}
\end{equation}

The numerator of the Reynolds number includes parameters related to inertia (density $\rho$, velocity $v$ and dimension $L$), and the denominator describes the viscosity $\eta$ of the fluid, which makes this a perfect parameter to understand movement in a fluid. It follows that for small Reynolds numbers ($\tx{Re} \leq 10^{-2}$) inertia effects are negligible in contrast to viscous forces. Specifically, an accelerated particle will reach its maximum speed very quickly and its movement will be halted almost instantaneously when the force stops acting. An example of a system with very small Reynolds numbers are bacteria moving in intracellular space. Moreover, in the range of low Reynolds numbers, fluid flows are usually considered laminar. 

In contrast, for large Reynolds numbers ($\tx{Re} > 20$) inertial forces are large compared to viscous forces. A particle  with big $L$ under the action of a force will need a certain time to reach a final velocity. Examples of this are aeroplanes or tankers, both needing some relaxation time to come to a standstill when the motors are switched off. In such regimes, inertia tends to deviate the fluid flows from well behaved and laminar to turbulent \cite{Fluid_Dynamics55}.

\subsection{Self-propelled particles}

In order to generate movements on small scales, i.e. in the low Reynolds number regime, one requires propulsion mechanisms that are different from the ones known at the macro-scale. Nature has evolved many such friction dominated propelling mechanisms like the spiral movement of flagella in some bacteria\cite{berg1973bacteria} or the paradigmatic Scallop movement\cite{Purcell_LifeLowReynoldsNumber}. 

Research to design particles that can autonomously move in such environments has led researchers to a very promising concept: the fabrication of self-propelled particles. These are particles that use various mechanisms to gather energy from their environment and convert it to movement. One example of this is \textit{electrophoresis}, in which two chemical reactions take place on the surface of the SPP, releasing different ions to the surrounding fluid. This is turn creates a local electric field through which the particle itself is transported \cite{electrophoresisRodsGold, electrophoresisRodsGoldExplanation, electrophoresisRodsCopper, villa2019fuel-free}. In \textit{diffusiophoresis}, a chemical reaction creates a concentration gradient around the SPP through different diffusion constants of various products of the reaction \cite{CatalysisEnzymeDiffusion, diffusiophoresisAgCl, waterFilterDiffusiophoresis, villa2019fuel-free}. Random collisions with the products accelerate the SPP due to the concentration gradient in a specific direction. \textit{Thermophoresis} also makes use of this movement generated by different amounts of diffusion on different sides of the SPP: Here, in contrast, the diffusion gradient is not generated by a chemical reaction, but by a temperature gradient on the particle surface. Janus particles with two different coatings for example can absorb heat anisotropically and generate such self-propulsion effect \cite{selfThermophoresis}.

\subsection{Light as a propulsion mechanism}

For some of those mechanisms, chemical reactions take place on the particle's surface with chemicals in the environment and accelerate the particles to speeds in the order of several hundreds of $\mathrm{\mu m/s}$  \cite{mou2015single-component}. Particular applications of the propulsion mechanisms mentioned above usually rely on the existence of a "fuel" in the particle environment, which reacts with the particle surface to induce a propulsion mechanism. Highly reactive molecules like $\mathrm{Br_2}$ \cite{electrophoresisRodsCopper} or $\mathrm{H_2O_2}$ \cite{electrophoresisRodsGold} are typically used, which prevents the usage of such technologies in biological systems, and requires a constant refuelling of the environment. 

The pursue for bio-compatible SPPs has led to the invention of original ways of propulsion such as magnetic actuation \cite{tottori2012magnetic, jahani2026fabrication} or even acoustic propulsion \cite{xu2017ultrasound,voß2021acoustically}. 
However, little research has been conducted on a particularly accessible, bio-compatible and easy to control energy source: light. 
While some proof-of-concept can be found in literature\cite{yang2023light-powered, sipovajungova2020nanoscale, AlejandroMasterThesis} a comprehensive study on particles propelled simply by light momentum transfer is still missing. In this work, we study this promising new approach both experimentally and in simulations.

A light-based propulsion system could offer a solution to the problem of biocompatibility, as here the energy for propulsion does not come from a chemical reaction with the environment, but from the interaction with light.

\subsection{Light as a propulsion mechanism}
\begin{figure}[h]
\centering
  \includegraphics[width=0.4\textwidth]{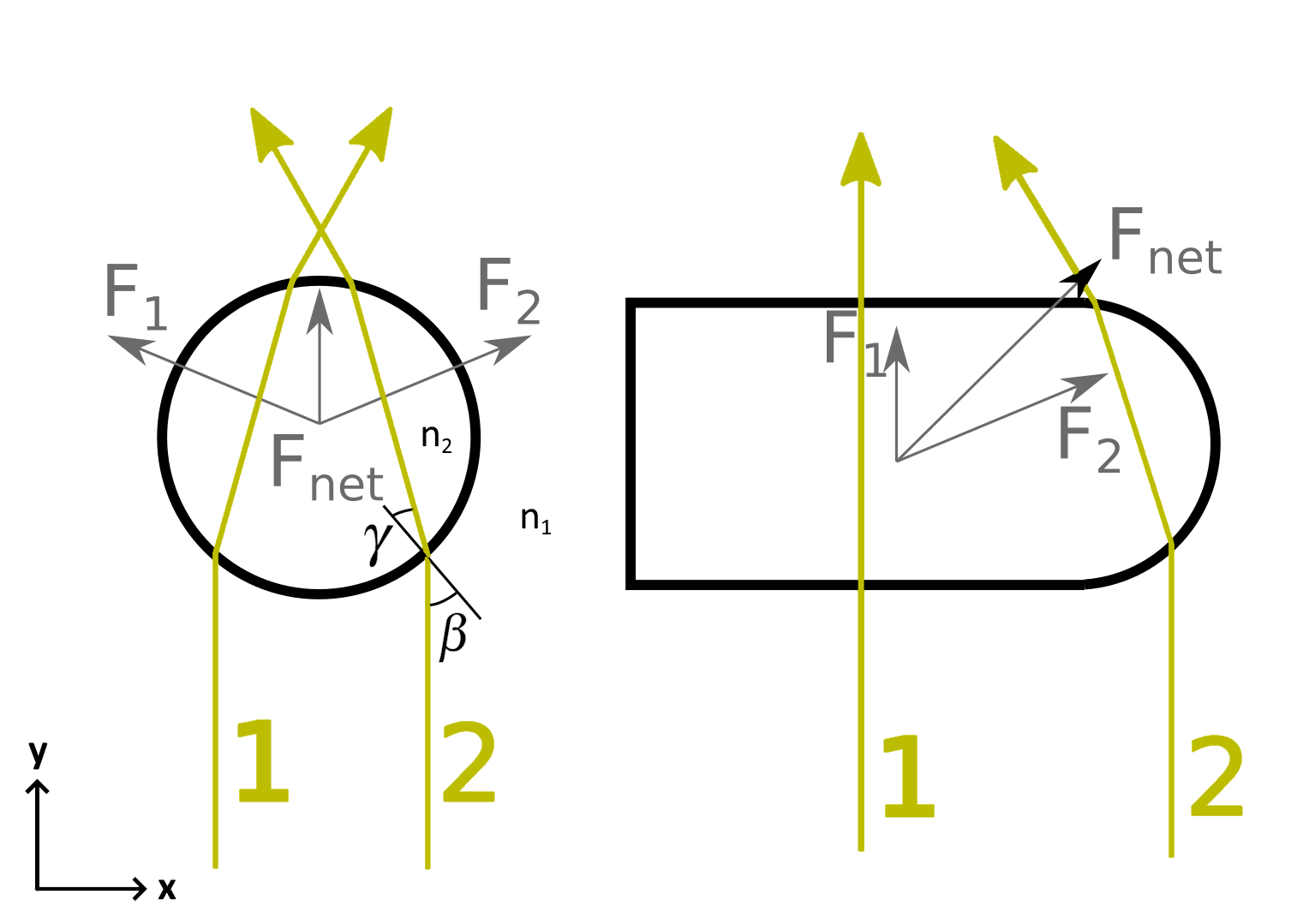}
  \caption{A sphere and a bullet shaped 2D particle with the corresponding optical forces resulting from the momentum transfer of the respective beams, as well as the resulting total force. The left particle represents a basic working principle of Optical Tweezers, in which a force equilibrium is reached when the refractions are symmetrical and the particle is effectively "trapped". By breaking the shape symmetry, a net force can be achieved, which can be harnessed for propulsion.
  All angles are spanned from the beam to the surface normal.}
  \label{fig:opticalForces}
\end{figure}

An experiment that defined a milestone in the history of microscopic manipulation was the application of light momentum transfer to build an optical trap by Ashkin in 1986 \cite{ashkin1986optical}. Its use is spread nowadays under the name of Optical Tweezers, and it goes to show how precise and powerful the action of light can be at the microscale. The same principle of light momentum transfer used in Optical Tweezers could be applied to the design of particles which preferentially deflect light in a particular way, enabling them to be light-propelled (see Figure \ref{fig:opticalForces}). Even though light is an easily controlled energy source, generally not harmful to biological tissue and considered very efficient for self-propulsion mechanisms, almost no research has been conducted on the creation of light-propelled microparticles up to date \cite{yang2023light-powered}. An example of the use of light as a drive mechanism was presented by Peterson et al. in 2010 with their stable optical lift. The authors used a refractive object with different upper and lower surfaces and irradiated it with a homogeneous laser beam. The transfer of momentum during the refraction of light at the objects surface therefore created a force that could lift the object \cite{peterson2010frontiers}. Other works in which light momentum transfer has been used as a propulsion mechanism have taken advantage of simple reflection \cite{buzas2012light} or the differential refraction of polarized light\cite{Liu2010light-driven}. Given the lack of further research into this promising propulsion mechanism, we aim to study in detail the conditions in which a particle can be propelled by the differential refraction of a homogeneous light source.\\

\subsection{Light momentum transfer}
When light reaches the interface between two media with different refractive indices its straight-line trajectory is deviated, i.e. the light beam is refracted. Given that the angle $\beta$ of incidence with respect to the surface normal $\vec{e_n}$ and both refractive indices $n_1$ and $n_2$ are known, Snell's Law allows the calculation of the new light angle $\gamma$ such that

\begin{equation}
    \frac{\sin{\beta}}{\sin{\gamma}} = \frac{n_2}{n_1} \tx{ .}
    \label{eq:snell}
\end{equation}

As long as the complete geometry of simple objects like the ones depicted in Fig. 1 can be fully expressed in a concise mathematical shape (rectangles, triangles, circles) the full refraction picture can be easily reconstructed as a first step towards the calculation of momentum transfer.
Each photon in a light beam carries a momentum $p$ given by 

\begin{equation}
    p=n\frac{h}{\lambda_0} \tx{ ,}
\end{equation} 

with $h$ the Plank's constant and $\lambda_0$ its wavelength in vacuum. A change in its trajectory implies a momentum change. A consequence of the trajectory deflection is thus a momentum transfer to the refractive object given by
\begin{equation}
    \Delta \vec{p}=\frac{h}{\lambda_0}(n_2 \cos\gamma - n_1 \cos\beta)\cdot \vec{e}_n.
    \label{eq:momentum_transfer}
\end{equation}
The reader might notice how the momentum transfer to a refractive object always happens parallel to the incidence interface normal, $\Delta\vec{p}\parallel\vec{e}_n$. Also, the momentum transfer of each photon can be increased by using smaller wavelengths and bigger refractive index difference between the two media.

\section{Materials and Methods}

\subsection{Experimental design}

The lack of literature about particles propelled by light refraction presented a challenge in the conceptualization of the proof-of-principle experiments. First, in order to test the feasibility of such propulsion mechanism we chose a combination of microfabrication with polymer printing technology and the design of a dedicated optical setup, as described in the following sections.\\

\subsubsection{Two-Photon Polymerization of particles}

In this work, Two Photon Polymerization (TPP), as first presented by Maruo et al., has been used for the fabrication of self-propelled microparticles  \cite{maruo1997Three-dimensional}. In contrast to previous methods like photolithography, TPP is based on the simultaneous absorption of two photons by a resin mesh to polymerize. The smaller cross section of this phenomenon increases dramatically the printing resolution in contrast to previous photolitographic technologies \cite{2PPResolution}. Our particles were produced printing the resin IP-Dip2 in the Photonic Professional GT2 (PPGT2, Nanoscribe GmbH \& Co., Germany), which in combination can reach a resolution between $400$ and $\mathrm{500\,\mu m}$ \cite{NanoScribeResolution, NanoScribeResolutionSupport}. 

To prove the feasibility of refractive-based propulsion and test the influence of particle geometry, spherical and bullet-shaped particles, as sketched in Figure \ref{fig:opticalForces}, were fabricated in 3D. The specific particles are chosen, as the sphere is fully symmetric, so it is expected that the resulting forces in direction orthogonal to a homogenous illumination get cancelled out, while a resulting force in the direction of the incident light remains. For the bullet, we expect that the broken symmetry leads to a net force, normal to the incident light, which makes this shape a good candidate to generate optical forces by homogenous illumination. The spheres have a radius of $\mathrm{5\,\mu m}$ and the bullets consists of a hemisphere with a radius of $\mathrm{5\,\mu m}$ at the front and a matching cylinder with a length of $\mathrm{20\,\mu m}$ at the back. The particles are printed on a fused silica substrate, which has a suitable refractive index for the printing. In short, the desired 3D objects were designed in \textit{Blender} and distributed in typical arrays of 20x20 particles, covering a total printing surface of about $0.16\,\mathrm{mm^2}$. Identical arrays of bullet and sphere particles were created and printed simultaneously onto the same silica substrate using a PPGT2 device through a 63x objective. The silica substrate was loaded with a single IP-Dip2 droplet in which the print took place. After the polymerization, the sample was developed and the remaining IP-Dip2 was washed using mr-Dev 600 (Micro Resist Technologies GmbH) for a duration of 10 minutes, before cleaning with acetone and performing a last drying cycle with $N_2$.

\subsubsection{Particle chamber}
\label{subsec:particleChamber}

Light-propulsion experiments are performed inside a custom chamber where our micrometer-sized particles can move freely within a confined space, which facilitates the localization and recording of movement. A piece of double-sided adhesive tape with a thickness of $d_\tx{tape} = 260\,\mathrm{\mu m}$ is used, where a hole with a radius of $r_\tx{hole} = \mathrm{1.5\,mm}$ is punched inside it. The adhesive tape is stuck to the silica substrate leaving the region with the printed particles inside the hole. Afterwards, the chamber is filled with distilled water with some overflow ($V_\tx{hole} = r_\tx{hole}^2 \pi d_\tx{tape} = \mathrm{1.84\,\mu l}$) to avoid the formation of air bubbles and then sealed with a glass cover slip. Excess water naturally flows into the small spaces between the adhesive tape and the two glass panes when these are pressed together. To avoid water evaporation, the chamber is kept hermetically sealed by closing the space between both glass slides with nail polish. A schematic of the particle chamber is shown in Figure \ref{fig:setup} A and B.
Before running an experiment, the entire particle chamber is placed in an ultrasonic bath for some seconds. The induced vibration from the bath causes the weak bond between the particles and the silica substrate to break, leaving the particles free floating without damaging them. To avoid that the particles spread through the entire volume or stick to the adhesive walls of the chamber, short pulses of 3 seconds are applied each time until most particles are detached. 

\subsubsection{Observation of light-propelled movement}

\begin{figure*}
 \centering
 \includegraphics[width = 0.588\textwidth]{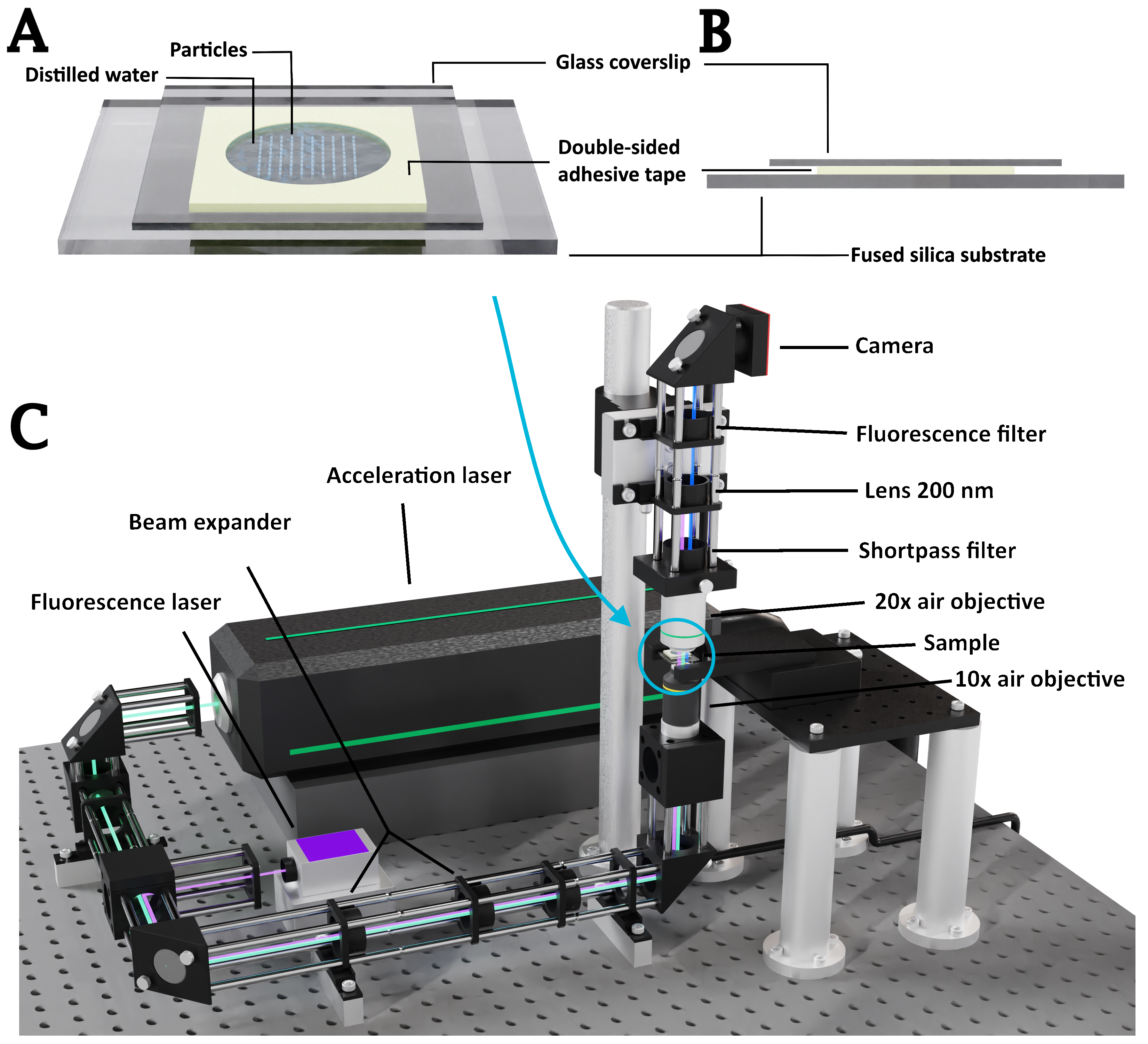}%width = 0.56\textwidth
 \caption{A: Structure of the home made sample chamber in which the printed particles are located. It consists of the fused silica substrate on which the particles are printed, double-sided adhesive tape with a hole with a radius of $\mathrm{1.5\,mm}$, distilled water and a glass cover slip. B: Side view of the sample chamber. C: The setup of the custom build microscope, consisting of the acceleration laser (green laser) and fluorescence laser (violet laser), various lenses and filters, a 10x air objective (yellow), a 20x air objective (green) and a camera. The fluorescence light detected by the camera is coloured in blue.}
 \label{fig:setup}
\end{figure*}

A custom setup was built in-house for the simultaneous propulsion and observation of the particles. This was achieved by integrating a $\mathrm{5\,W}$ solid state laser with a wavelength of $\mathrm{532\,nm}$ (Verdi, Coherent) propulsion laser via a magnification arm in a fluorescence microscope. Autofluorescence of the particles was induced by a $\mathrm{405\,nm}$ laser with a maximum intensity of $\mathrm{50\,mW}$ (Obis, Coherent). Both, the fluorescence and the propulsion laser were coupled in the back aperture of the illumination objective (10x, NA = 0.25). This is done by a beam expander (consisting of two lenses $\mathrm{25\,mm}$ and $\mathrm{250\,mm}$ focal length) and a lens (with a focal length of $\mathrm{100\,mm}$) to achieve homogeneous illumination with parallel light in the sample plane. Using a 20x, NA=0.70 air objective and a $\mathrm{200\,mm}$ tube lens, the image of the sample plane was recorded by a CMOS colour camera (CS165CU, Thorlabs) with a resolution of 1440 x 1080 pixels. To avoid the propulsion laser beam to enter the camera and outshine the desired autofluorescence signal from the particles, a shortpass filter with a cut-off wavelength of $\mathrm{500\,nm}$ (FESH0500, Thorlabs) is introduced after the imaging objective. A notch filter for $\mathrm{405\,nm}$ was additionally placed before the camera to remove the fluorescence-inducing laser. This leaves a clear fluorescence image of the particles to be recorded, which are auto-fluorescent between $\mathrm{400\,nm\ and\ 700\,nm}$ \cite{IP-DIP2}. The beam path of the fluorescence (green) and propulsion laser (violet) can be seen in
%turquoise ??
Figure \ref{fig:setup} C. 

Particle movement was recorded as a video with a frame rate of 2 fps, which was sufficient to capture the trajectories. 

\subsubsection{Analysis of particle movement}
Some steps of post-processing were necessary for the videos before the data could be analysed. From the three RGB channels available form the acquisition, only the blue one containing the particle fluorescence was extracted. The values for the grey levels in the image are then inverted and the movement analysis is done using the TrackMate plugin in FIJI \cite{tinevez2017TrackMate}. As a result, the trajectories of the particles are obtained visually superimposed on the input data (Figure \ref{fig:MSDFit1} A), as well as the coordinates of the particles over time. This position data can then be used to determine the mean squared displacement (MSD) of each trajectory. It is assumed that the particles move by normal and anomalous diffusion, which can be described in two dimensions by \cite{MSDAnalysis}

\begin{equation}
    \tx{MSD} = 4 D \left(\frac{\Delta t}{t_0}\right)^\alpha \tx{ ,}
    \label{eq:MSDFit}
\end{equation}

where $D$ is the diffusion coefficient, $\Delta t$ is a given time interval of the trajectory and the timescale is normalized by $t_0$, which is set to $\mathrm{1\,s}$. The exponent $\alpha$ distinguishes between different movement types: 
\begin{itemize}
    \item  $\alpha=1$: Particles exclusively undergo Brownian motion and the MSD is simply proportional to the time interval $\Delta t$. 
    \item $\alpha < 1$: Particle movement is subdiffusive, which can be observed in creep movements or diffusion in very crowded environments. 
    \item  $1<\alpha<2$: Particles movement is superdiffusive, which is for example typical in active cellular transport. 
    \item In all other cases, the movement is either constant in a given direction (ballisitic, $\alpha = 2$) or even hyperballistic ($\alpha > 2$) such as in an accelerating car. Examples of such movement have been generated in optical systems \cite{BeyondBallistic}. 
\end{itemize}
To determine which type of motion a particle follows, the MSD curves are fitted to \eqref{eq:MSDFit} and the exponent $\alpha$ is determined. 
During the development and detachment steps of the fabrication, some particles remain stuck to the silica substrate and do not move during experiments. We removed such outliers and restricted our analysis  to particles which presented a total displacement larger than $\mathrm{5 \,\mu m}$ (radius of the sphere) during the 10-minute recording. 

\subsection{Momentum transfer simulations}

After the experiments were performed, we generated simulations of expected optical forces to compare these with the experimental results and explore the influence of different geometries on refractive propulsion. Different 2D particles were generated as a combination of simple geometrical  shapes (see Fig. \ref{fig:rayDirection}) which allowed a straightforward analytical solution of the light refraction at each position. Combining \eqref{eq:snell} and \eqref{eq:momentum_transfer}, the momentum transfer of a photon bundle sharing the same trajectory can be computed.

However, a further consideration is necessary. At each interface, part of the light is not refracted but reflected with an angle $-\beta$ (cf. \eqref{eq:snell}). The proportion of refracted and reflected light can be described using the transmission and reflection coefficients $R$ and $T$, for which $R+T=1$ holds. In this work, Schlick's approximation is used, which provides a simple and fast method for calculating the coefficients \cite{schlickApproximation}:
\begin{align}
    \label{eq:schlickApproximation}
    R &= R_0 + (1- R_0)(1-\cos{\beta})^5\\
    \label{eq:tansmissionReflection}
    T &= 1 - R \tx{ ,}
\end{align}
where  $R_0$ is the value of the reflection coefficient for a light beam that falls perpendicular to the media interphase, given by

\begin{equation}
    R_0 = \left(\frac{n_1-n_2}{n_1+n_2}\right) ^2 \tx{ .}
\end{equation}

In the case of a transition from an optically denser to an optically thinner medium, total reflection ($R=1,\ T=0$) can occur at certain angles of incidence. In that case the momentum transfer is given by
\begin{equation}
    \Delta \vec{p} = 2 \frac{h}{\lambda_0} n_1 \cos\beta \cdot \vec{e_n} \tx{ .}
    \label{eq:momentumFromAngleTotalReflection}
\end{equation}
Once again, momentum transfer occurs perpendicular to the refraction surface (see \eqref{eq:momentum_transfer}). Once the momentum transfer of single photon trajectory has been calculated using \eqref{eq:momentum_transfer} and \eqref{eq:momentumFromAngleTotalReflection}, and given the light intensity is known, an integration over the whole illumination surface can be performed to obtain the total momentum transfer of the light beam on an object. The total force acting on the particle (see Figure \ref{fig:opticalForces}) can be determined by the sum of the momentum transfers per unit time.

\subsection{Analytical solution of the refraction}

\begin{figure*}[htbp!]
 \centering
 \includegraphics[width = 0.95\textwidth]{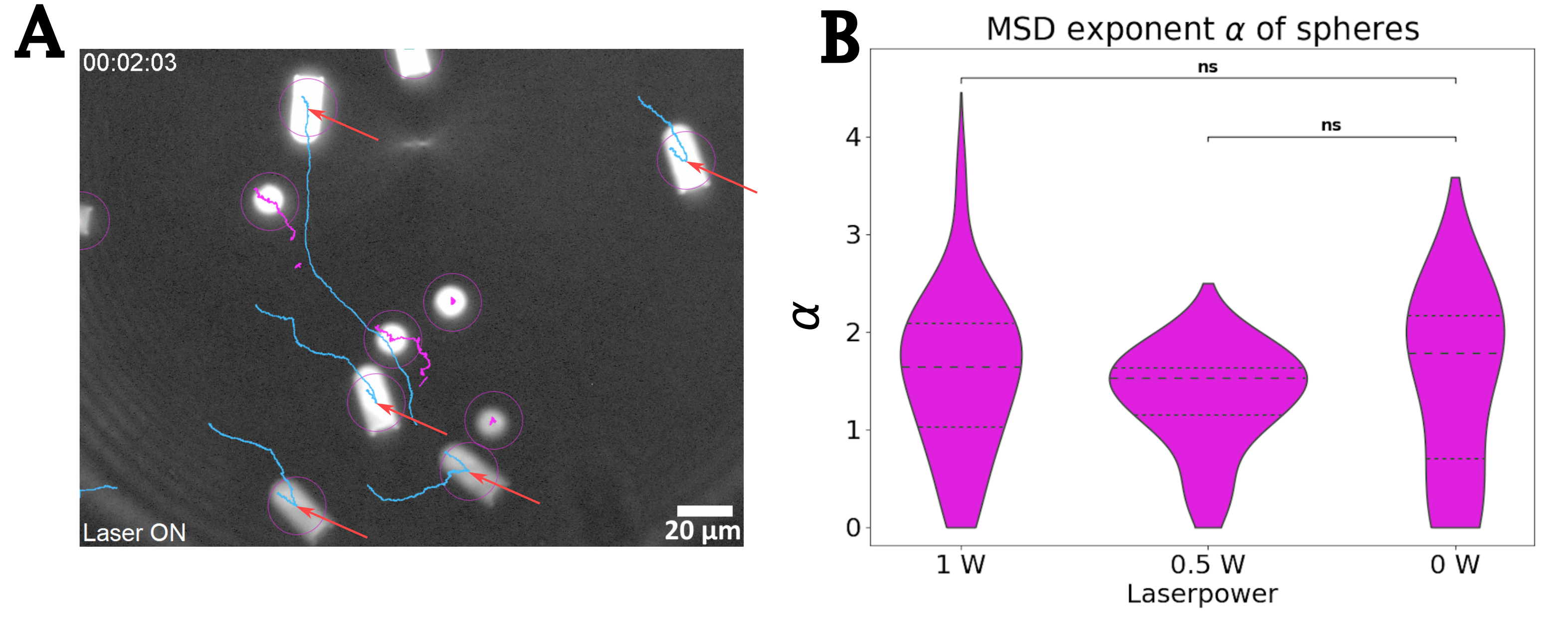}
 \caption{A: Trajectories of bullets (blue) and spheres (magenta). In the experiment, the accelerating laser had an intensity of $\mathrm{1\,W}$ and was switched on after two minutes, the time of the snapshot. The kink in the trajectories of the bullets corresponds to the moment the acceleration laser was turned on which led to a change in direction. The exact position of the bullets at this time point is marked by a red arrow. Supplementary Movie 1 shows the transition from pure Brownian motion to light-propelled movement. B: The MSD exponent for spheres for 36 trajectories at $\mathrm{1\,W}$, 7 trajectories at $\mathrm{0.5\,W}$ and 6 trajectories at $\mathrm{0\,W}$, determined by fitting the MSD of the trajectories to \eqref{eq:MSDFit}. The exponent with the laser switched on is not significantly different from the laser switched off (p \textgreater \  0.05, t-test for $\mathrm{1\,W}$ and $\mathrm{0.5\,W}$).}

 \label{fig:MSDFit1}
\end{figure*}

Simulations of the total refraction momentum transfer were performed in the object-oriented programming language Python. Several different particle geometries were investigated, in all cases using surfaces that could be mathematically described in closed forms. This facilitated the calculation of two Points of Impact (POI) per photon (when entering and when exiting the particle) without the necessity for intermediate steps. The only exception were photons which experienced total reflection for which only one POI was necessary (never entered the particle) or more than two POIs (experienced several refractions or reflections inside the geometry).

The combination of \eqref{eq:snell} and \eqref{eq:momentum_transfer} suffices for the complete solution of the problem without cumbersome  numerical methods, which increases dramatically the speed of the simulation. For full comparison, realistic experimental parameters were used in the simulation. Refractive indices of the medium ($n_{water} = 1.333$) and the particles ($n_{IPDip2} = 1.547$ according to the manufacturer) were used \cite{IP-DIP2}. Also, particle dimensions were kept constant and the experimental photon wavelength $\lambda_0 = \mathrm{532\,nm}$ was used. 

In a full ray-tracing simulation, 10.000 rays were propagated along the y-axis from negative to positive values (bottom to top in Figure \ref{fig:rayDirection}), with the same incidence direction as in the experiments. To have full coverage, rays were emitted at random positions across a plane below the particle, spanning its entire projected area. Care was taken to ensure that enough rays were calculated so that stochastic imbalances due to the random distribution could be neglected. In the simulation, each ray represents a real photon bundle which share the same trajectory. When the laser intensity and the number of simulated rays (photon bundles) were fixed in the simulation, the number of photons per bundle could be calculated. For each photon bundle, the fraction of reflected and refracted light was calculated with the aforementioned Schlick's coefficients R, T, with the special case of total reflection, R=1 (see \eqref{eq:schlickApproximation}).

Our analytical approach offers very fast solution times for each illuminated particle, since only one calculation per refraction or reflection is necessary, independent of the distance between the start of the beam and the POI. This approach is however only possible if the particles are composed of simple geometric figures which can be described in closed mathematical forms like rectangles, circles and triangles, as is the case here.

\section{Results and discussion}
\subsection{SPP driven by light}

A typical experimental dataset is shown in Figure \ref{fig:MSDFit1}A, where 10 minute trajectories are overlaid in magenta (spheres) and blue (bullets). In this acquisition, 2 minutes were recorded under no illumination before turning on the laser, which gave the chance to study and compare a typical, non-driven Brownian motion. Figure \ref{fig:MSDFit1}A and Supplementary Movie 1 captures the exact moment the laser was turned on. 
Already at first glance longer trajectories for the bullet particles can be noticed, which indicates how they moved faster in the same experimental time. It is also noticeable that four of the five bullet trajectories show a clear change in direction, with the moment of the change in direction corresponding to the moment the laser was activated. The exact position of the bullets at this time point is marked by a red arrow. This phenomenon can be explained by a small global drift in the sample, which is supported by the finding that all particles initially move towards the same direction (lower-right). For the bullet particles, the motion induced by the laser overcomes the global drift, leading to a new movement direction corresponding to their own orientation. 

In a first control quantification, we focus our attention onto the spherical particles. Since those particles are perfectly symmetrical, an a-priori analysis based on our momentum equations \eqref{eq:snell} and \eqref{eq:momentum_transfer}, indicates that no directed movement should arise under laser illumination. For this quantification, we ask whether illumination power would change the nature of the movement, as expressed by the exponent of the particles' MSD (see \eqref{eq:MSDFit}). Figure \ref{fig:MSDFit1}B shows the distribution of exponents $\alpha$ as measured for all spherical particles at $\mathrm{1\, W}$, $\mathrm{0.5 \, W}$ and laser OFF conditions ($\mathrm{0\,W}$). As expected for this particle geometry, illumination does not render different modes of motion and in all cases $\alpha$ presents a mean value below 2. The deviation from a simple Brownian motion ($\alpha=1$) is well explained by a slow liquid flow leading to the aforementioned drift, which can also be observed in Figure \ref{fig:MSDFit1} A (magenta trajectories).  

\subsubsection{Quantification of the propulsion}

\begin{figure*}[htbp!]
 \centering
 \includegraphics[width = 0.95\textwidth]{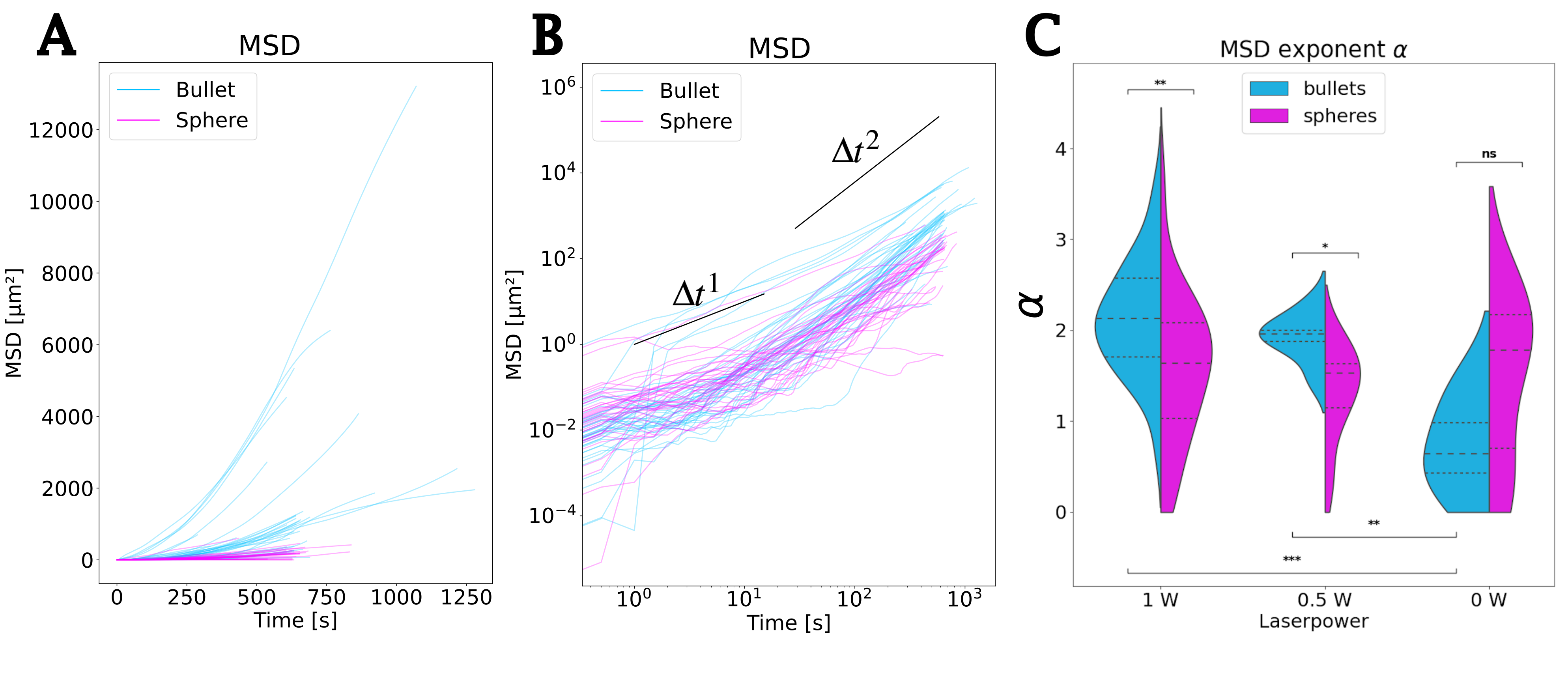}
 \caption{A, B: MSD curves determined from particle trajectories of spheres and bullets at an accelerating laser intensity of $\mathrm{1\,W}$. The data is plotted with linear axes (A) and as a log-log plot (B). In the latter, a straight line with a slop of one ($\Delta t^1$) and a parabola ($\Delta t^2$), recognisable in the log-log plot as a straight line with a slope of 2, were superimposed as a guide to the eye. C: The MSD exponent $\alpha$, determined by fitting the MSD of the trajectories to \eqref{eq:MSDFit}, for bullets (blue) and spheres (magenta) at different laser powers. For a laser power of $\mathrm{1\,W}$ and $\mathrm{0.5\,W}$, $\alpha$ is significantly higher for the bullets compared to the spheres (p \textless \  0.01, t-test for $\mathrm{1\,W}$ and p \textless \  0.05, t-test for $\mathrm{0.5\,W}$). In contrast, there is no significant difference between bullets and spheres when the laser is turned off ($p \simeq \  0.26$, t-test).
 For the bullets, the exponent with the laser switched on is significantly higher than with the laser switched off (p \textless \  0.001, t-test for $\mathrm{1\,W}$ and p \textless \  0.01, t-test for $\mathrm{0.5\,W}$). There are 76, 13 and 9 trajectories analysed for a laser power of $\mathrm{1\,W}$, $\mathrm{0.5\,W}$ and $\mathrm{0\,W}$.}
 \label{fig:MSDFit2}
\end{figure*}

A total of 37 experiments were performed in identical conditions as in the previous section to create a meaningful dataset with enough movement statistics. After particle tracking and data pooling, the MSDs of sphere and bullet particles were computed and plotted as shown in Figure \ref{fig:MSDFit2} A. In all cases, bullet particles exhibit higher values of MSD for the same time intervals, which reveals a more long-range, directed movement. 

Figure \ref{fig:MSDFit2} B shows the same data in log-log space, where the slope of the curves directly reveals the exponent $\alpha$ of the MSD (see \eqref{eq:MSDFit}).
Here, both types of particles exhibit a flatter MSD with lower $\alpha$ for time intervals between 1 and 10 seconds. At those time scales, movement seems to be still dominated by pure Brownian motion, for which an ideal proportionality with $\Delta t ^1$ should apply. 
For longer time intervals ($\Delta t > \mathrm{10\,s}$), the MSD of bullet-shaped particles becomes more steep (see Figure \ref{fig:MSDFit2} B), with an average slope which seems to get closer to values of $\alpha$ related to directed motion and propulsion, $\alpha\geq2$.

Figure \ref{fig:MSDFit2} C shows a direct comparison of motion exponent $\alpha$ between bullet-shaped particles and spheres for different laser powers. For each illumination assay ($\mathrm{1 \, W ,\ 0.5\, W } $), bullet particles exhibit a greater and significantly different value of $\alpha$ compared of the spheres, indicating a more directed motion within the same experimental conditions. In the case where the sample was not illuminated ($\mathrm{0\,W}$), no significant difference can be observed between the motion of bullets and spheres. This supports the expectation  that the particle geometry plays a crucial role in the creation of a self-propelling refraction of bullet shapes. Furthermore, when comparing bullet population among themselves (Figure \ref{fig:MSDFit1} C, bottom significance markers), statistical significance between the different laser power situations is found. All these results combined indicate that bullet particles are able to propel themselves by means of engineered refraction, and that the laser power used for homogeneous illumination has a direct impact on their directed motion speed. 

\subsubsection{Determining the speed of the bullets.}

By fitting the MSDs to \eqref{eq:MSDFit}, not only the exponent $\alpha$, but also the prefactor $D$ can be determined. The mean value of this prefactor for the bullets at $\mathrm{1\,W}$ and for the bullets at $\mathrm{0.5\,W}$ are:

\begin{align*}
    \bar{D}_{\mathrm{1\,W}}   &= \mathrm{(1.04\pm0.40) \cdot10^{-2}\,\mu m^2}\\
    \bar{D}_{\mathrm{0.5\,W}} &= \mathrm{(0.31\pm0.25)\cdot 10^{-2}\,\mu m^2} \tx{ .}
\end{align*}

For a constant movement with velocity $v$ (which means that $\alpha$ is assumed to be 2), the MSD can be described by: $\tx{MSD} = v^2t^2$. By comparing this with \eqref{eq:MSDFit}, the relationship $v = \frac{\sqrt{4 D}}{t_0}$ can be derived. With this, the velocities can be calculated from the prefactor $D$:

\begin{align*}
    v_{\mathrm{1\,W}}   & = \mathrm{0.20\pm0.04\,\mu m/s}\\
    v_{\mathrm{0.5\,W}} & = \mathrm{0.11\pm0.05\,\mu m/s} \tx{ .}
\end{align*}

This result is a further confirmation of the influence of illumination on self-propulsion. While the previous result from Figure \ref{fig:MSDFit2}C hinted to an increased directed motion for higher laser powers, the velocity calculation reveals a linear proportionality between propulsion speed and photon count and thus laser intensity, as expected theoretically. 

\subsection{Simulation: Momentum transfer of light}

\begin{figure*}[htbp!]
 \centering
 \includegraphics[width=0.99\textwidth]{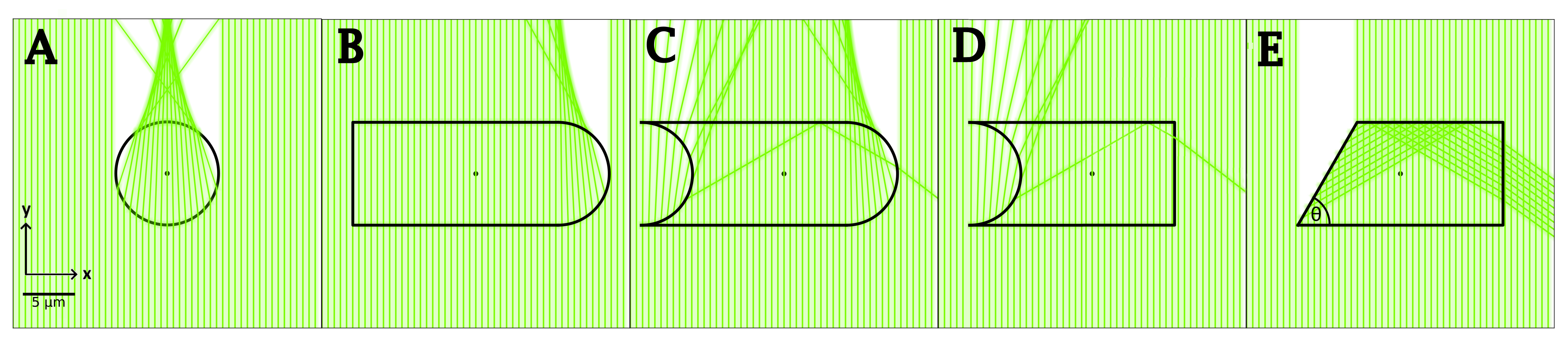}
 \caption{Refraction of light rays for all five investigated shapes, from left to right: sphere, bullet, double bullet, reversed bullet and right trapezoid. For the latter, $\theta$ is the slope-angle for the total reflection.}
 \label{fig:rayDirection}
\end{figure*}

Our experiments with spheres and bullet particles confirm the prediction that particles can be propelled by light refraction even under homogenous illumination. As expected, upon laser illumination, the movement is significantly different, and an increase in laser power (i.e., more photons being refracted) noticeably boosts particle movement in a linear manner.
Since the main result of our experiments revealed that shape-symmetry breaking is the key feature for refraction-propulsion, we developed a ray-tracing simulation to further investigate the influence of particle shape.

Since a simple reflective surface can generate a momentum transfer parallel to the beam direction very easily, the focus here is particularly on the x-direction, i.e. on the momentum transfer orthogonal to the beam direction $\Delta p_x$ (cf. Figure \ref{fig:opticalForces}). The particles to be analysed can be found in Figure \ref{fig:rayDirection}. 
From left to right, these are the sphere, the bullet, the double bullet, the reverse bullet and the right trapezoid. Due to their different shapes, the five forms described above each have a different influence on the direction of light refraction and therefore also on the momentum transfer.

\subsubsection{Exploration of geometry}

Figure \ref{fig:rayDirection} A shows the full refraction picture of a spherical particle when illuminated with a homogeneous beam of light from below. Here, each vertical ray represents a photon batch sharing the same direction and travelling from along the y direction. As can be observed, the symmetry of the particle itself prohibits any final contribution to lateral x-momentum, since light rays are refracted symmetrically around the sphere's centre. Instead, the only remaining contribution is along the vertical y-direction, where the beams loose a part of their momentum which is transferred to the particle.

Since a movement perpendicular to the illumination direction is desired, the symmetry of the particle along the y-axis must be broken. As well as in our first experiments, we design a bullet-shaped particle: a cylinder with a hemisphere attached to one end. Figure \ref{fig:rayDirection} B shows its refraction effect: while the cylindrical body does not contribute to the momentum transfer, the spherical cap deflects light towards the left (flat) side of the bullet, creating then a net momentum transfer towards the curved end (right).

\subsubsection{Total internal reflection as momentum transfer boost}

Since the bullet-shaped particle clearly demonstrates that a disruption of particle symmetry leads to a net momentum transfer, we explore further asymmetric shapes in our simulations. In our first iteration, we add a concave spherical cup at the rear of the bullet particle, creating a "double bullet" (Figure \ref{fig:rayDirection} C). The ray-tracing simulation renders a surprising result at first glance: not only the magnitude of the momentum transfer has diminished, but its direction has also been reversed. Careful observation of the ray tracing reveals the reason behind this result: The convex part of the bullet refracts light as previously seen in the bullet, but the concave end creates a refraction mostly pointed in the opposite direction, counteracting the net x-momentum transfer of the bullet head. For the sign of $\Delta p_x$ to have changed, this new momentum transfer contribution must be even stronger than in the bullet alone, and the explanation can be again recognized in the ray tracing: photon bundles experience total internal reflection, which contribute massively to $\Delta p_x$, as those rays have a stronger change in direction.\\
This result leads us to consider a new geometry with a single concave element, the "reversed bullet" (Figure \ref{fig:rayDirection} D). As expected, the net momentum transfer for this geometry is even greater than for the single bullet and also than the double bullet, now that no counteracting refraction elements are present. A decisive advantage of the reverse bullet over the previous iterations is its capacity to generate total internal reflection, greatly contributing to momentum transfer perpendicular to the illumination direction. To maximize this effect, we propose a shape with a side in which total internal reflection always happens. This corresponds to a right trapezoid in which one side always presents an angle above the critical angle for total reflection with respect to the beam. The critical angle can be calculated using $\theta_\tx{tot} = \arcsin{\left(\frac{n_\tx{medium}}{n_\tx{particle}}\right)}$, and is only depending on the refractive indices of the medium and the particle. For our materials, the value computes to
\begin{equation}
    \theta_\tx{tot} \approx 59.28^\circ { .}    
\end{equation}
To always ensure total reflection, our angle is set to $\theta = 60\,^\circ$ as seen in Figure \ref{fig:rayDirection} E.\

The values of net momentum transfer as obtained from our ray tracing simulations can be compared in Table \ref{tab:absoluteMomentumTransfer}. As can be seen from the above considerations, the sphere only has a very small x-momentum transfer arising as an artifact of the random spawning positions of the simulated rays. In comparison, the bullet has a higher x-momentum transfer arising from its asymmetric shape and is used as a reference. In the double bullet, the two ends work against each other, whereby the inward-facing part of the double bullet has a stronger contribution. This is isolated in the reverse bullet, which is why its x-momentum transfer is the greatest compared to the other bullet variants. The largest x-momentum transfer can be found in the right trapezoid, where the effect of total reflection is utilised to the greatest extent. For this particle the x-momentum transfer is more than 5 times greater compared to the bullet used in the experiments, rendering it as the best candidate for future experiments.

\begin{table}[h]
\small
  \caption{Absolute value of momentum transfer in x-direction $|\Delta p_x|$ for all five basic forms of the particle normalized to the momentum transfer of the bullet.}
  \label{tab:absoluteMomentumTransfer}
  \begin{tabular*}{0.48\textwidth}{@{\extracolsep{\fill}}ll}
    \hline
    Particle & $|\Delta p_x|$ \\
    \hline
    sphere & 0.001\\%0.004 \cdot 10^-23 [a.u.]\\
    bullet & 1\\%4.497 \cdot 10^-23 [a.u.]\\
    double bullet & 0.540\\%2.428 \cdot 10^-23 [a.u.]\\
    reversed bullet & 1.712\\%7.697 \cdot 10^-23 [a.u.]\\
    right trapezoid & 5.267\\%23.686 \cdot 10^-23 [a.u.]\\
    \hline
  \end{tabular*}
\end{table}

\section*{Conclusions}

In this work, we designed, manufactured and tested particles that can be propelled simply by the refraction of a homogeneous light beam, demonstrating the feasibility of the propulsion principle. In a further step, we explored the influence of geometry on the optimization of refraction through several iterations in a ray tracing simulation, leading to an optimized strategy based on total internal reflection.

In the experimental part, the movement of bullets and spheres was compared using different metrics, whereby a significantly faster and more directed movement was observed for the bullets for different laser intensities. This results demonstrates the effect of directed refraction caused by a geometrical symmetry breaking, which can be used as a design principle. In our simulations, large differences in momentum transfer orthogonal to the illuminating beam were found among the different shapes, where the right trapezoid was found to have the largest momentum transfer caused mainly by total internal reflection. 

Overall, our work is a proof-of-principle for an affordable, easy to implement and highly controllable mechanism for self-propulsion at the micro-scale, based only on the tailored refraction of light in asymmetric particles. Our fuel-free approach might solve one of the most common issues for the use for SPPs in biological environments, namely biocompatibility. A collimated, non-focused beam of light is enough to generate movement without noticeable changes in local temperature, a further advantage for in-vivo applications. In summary, together with acoustically propelled particles \cite{voß2021acoustically}, light propelled particles present the most promising alternatives for biocompatible propulsion at the micro-scale, for applications where chemical- or thermal- based propulsion systems might be harmful. 

\section*{Credit authorship contribution statement}

Jannis: Data Curation, Formal Analysis, Investigation, Methodology, Software, Visualization, Writing - Original Draft\\
\nd Alejandro: Methodology, Supervision, Writing – review \& editing\\
\nd Timo: Conceptualization, Funding Acquisition, Resources, Supervision, Writing – review \& editing

\section*{Conflicts of interest}
There are no conflicts to declare.

\section*{Data availability}

Data will be made available on request.

\section*{Acknowledgements}

This work was supported by the Deutsche Forschungsgemeinschaft (DFG, German Research Foundation): Project-ID 449750155 – RTG 2756, Project B3, Project-IDs 516046415; 456112451 and under Germany’s Excellence Strategy (EXC 2067/1- 390729940).

\balance

%%%REFERENCES%%%
\bibliography{references}
\bibliographystyle{rsc}
\end{document}